# THE TITIUS–BODE LAW OF PLANETARY DISTANCES: NEW APPROACH.

V. M. Bakulev

*Institute of Physics, St.Petersburg State University, Ulyanovskaya 1, Petergof, 198504 St.Petersburg, Russia;* vmbak@pobox.spbu.ru

**The striking regularity in the distribution of planetary distances has been the object of much attention for about two centuries. The first phenomenological formulation of the dependence between the numbers of the planets and their mean distances from the sun was proposed by Titius (1766), as:**

$$R_n = 4 + 3 \cdot 2^n \qquad (1)$$

**Where the radius of the earth's orbit is taken as 10, *n* is -∞ for Mercury, and 0,1,2, ... for succeeding planets and for Asteroid Belt between Mars and Jupiter. Ever since, this equation ("the Titius-Bode law") suffered numerous modifications (the history of the point is entirely described by Nieto [1]). Hypotheses of the solar system evolution considered predominantly the Titius-Bode law to provide some theoretical grounding, in spite of its notable gap between predicted and actual planetary distances for Neptune and Pluto. There is much effort of numerical simulation to justify its possible physical significance and to arrive at a higher approximation (references in [1-3]). Recently Nottale [4,5] has obtained quantification of orbits different from geometrical progression in terms of fractal trajectories governed by Schrodinger-like equation. The orbits of the planets are then equivalent to the Bohr circular orbits. However, in order to agree predictions of the theory with observations, Nottale postulates two not occupied orbits between** **the Sun and Mercury ($n=3$ for Mercury) and divides the solar system in two subsystems with a different normalizing parameter. Here I present more straightforward way of looking at the regular spacing of planetary orbits and show that the relative planetary distances can be represented as the inverse probabilities composed of two well-known distributions with one fitted parameter. It is likely similar to the Bose-Einstein condensation of degenerate gas in gravitational field.**

Three main characteristics of planetary motion: semi-major axis ($a_n$), angular momentum ($L_n$) and total energy ($E_n$) of the planet on $n$-orbit are related to each other by two equations:

$$L_n = m_n \sqrt{GM_0 a_n (1 - e_n^2)} \qquad E_n = -\frac{GM_0 m_n}{2a_n} \qquad (2)$$

Where $G = 6.672 \cdot 10^{-11} m^3 / kg \cdot \sec^2$ is gravitational constant, $M_0 = 1.989 \cdot 10^{30} kg$ is the mass of the Sun, $m_n$ is the mass of $n$-planet and $e_n$ is $n$-orbit's eccentricity.

Therefore, if the dependence of any one of these characteristics has integer representation, it follows that the same has to be valid for two others. The crude analogy between the solar system and atomic structure makes it tempting to believe that these representations can be formulated in terms of Bohr atomic orbits. This hypothesis will be more justified if angular momentum and energy are expressed in terms of quantities per unit mass. The data are shown in Table 1, where the mass of hydrogen atom $m_H = 1.673 \cdot 10^{-27}$ kg taken as unit mass is inserted in Eq. (2) instead of $m_n$ and orbital parameters of the most massive small planet in Asteroid Belt –Ceres are taken for the fifth orbit. In this case the dependence of $L_n$ upon $n$ reveals the notable regularity, the extremes of which fall on a straight line that passes through the origin of the coordinates and has the slope $\Delta L = 4.54 \cdot 10^{-12} J \cdot \sec$ (Fig.1). The resemblance to the Bohr model of hydrogen atom is appeared to be evident, if some reasons being responsible for the conspicuous deviation from the straight line in the middle part of the sequence. If it is true, then the dependence of energy per unit mass also can be expressed as a sum of two functions: one is proportional to $1/n^2$ and another one must be an asymmetric bell-shaped function. A number of simple one- or two-parametric functions can be expected to match this requirement, but computer fitting preferred the following: $n^3/(\exp(b*n)-1)$. Thus, the energy distribution was represented by:

$$E_n = A\left(\frac{1}{n^2}\right) + B\left(\frac{n^3}{\exp(bn)-1}\right), \quad (3)$$

Weight coefficients $A$, $B$ and exponent index $b$ were evaluated by fitting data in Table 1 (6 column) as:

$$A = -(1.507 \pm 0.058) \cdot 10^{-18} J \quad B = -(1.006 \pm 0.065) \cdot 10^{-18} J \quad b = 1.283 \pm 0.014$$

Hence in Eq. (3) $E_n$ constitutes the two-term sum of two convergent numerical series with corresponding weight coefficients. Every coefficient is appeared to be inversely proportional to the sum of corresponding series. This notice allows rearranging Eq. (3) within the same accuracy to give the reduction of fitted parameters in number. Taking into account that:

$$\sum_{n=1}^{\infty}\left(\frac{1}{n^2}\right) = \frac{\pi^2}{6} \qquad \sum_{n=1}^{\infty}\left(\frac{n^3}{\exp(bn)-1}\right) \approx \int_0^{\infty} \frac{x^3 dx}{\exp(bx)-1} = \frac{\pi^4}{15b^4}$$

we have:

$$E_n = \frac{E}{2}\left[\frac{6}{\pi^2} \times \frac{1}{n^2} + \frac{15b^4}{\pi^4} \times \frac{n^3}{\exp(bn)-1}\right] \quad (4)$$

Where already two not three variables are evaluated as:

$$E = -(4.88 \pm 0.09) \cdot 10^{-18} J \quad b = 1,283 \pm 0,013$$

The calculation results by Eq. (4) are presented in Fig. 2 and Table 1 (7 column).

The value of $E$ closely matches the sum of ten experimental energy quantities in Table 1 (6 column), that is equal to $-4.745 \cdot 10^{-18} J$. Then $E$ is most likely to be a sum of $E_n$ while $n$ tends to infinity:

$$E = \sum_{n=1}^{\infty} E_n = -(4.88 \pm 0.09) \cdot 10^{-18} J$$

In this case the ratio $E_n/E$ might be considered as a probability. If $E$ is rearranged from the right side of Eq. (4) to its left side then the remainder in the right side must be a probability too:

$$W_n = \frac{E_n}{\sum_{n=1}^{\infty} E_n} = \frac{1}{2}\left[\frac{6}{\pi^2} \times \frac{1}{n^2} + \frac{15b^4}{\pi^4} \times \frac{n^3}{\exp(bn)-1}\right] = \frac{1}{2}(W_{n1} + W_{n2}) \qquad (5)$$

Where:

$$W_{n1} = \frac{1/n^2}{\sum_{n=1}^{\infty} 1/n^2} = \frac{6}{\pi^2} \times \frac{1}{n^2} \qquad W_{n2} = \frac{n^3/\exp(1.283n)-1}{\sum_{n=1}^{\infty} n^3/\exp(1.283n)-1} = \frac{15b^4}{\pi^4} \times \frac{n^3}{\exp(bn)-1}$$

Indeed, according to the probability theory [7], the Eq. (5) might be a composite probability, if both distributions make up the full set of events. In this case the right side of Eq. (5) is the equal probability of two incompatible events: whether the occurrence of "$n$" in the distribution $W_1$, or the occurrence of "$n$" in the distribution $W_2$.

The expression for $a_n$ can be derived easily from Eq. (2) and Eq. (5):

$$a_n = a_0(W_n)^{-1} = 2a_0(W_{n1} + W_{n2})^{-1} \qquad (6)$$

Where $a_0 = -GM_0 m_H / 2E = 2.274 \cdot 10^{10} m$ provided $m_H = 1.673 \cdot 10^{-27} kg$.

It can be seen from Eq. (6) that the relative planetary distances $a_n / a_0$ do not depend on the unit mass quantity and can be presented as the inverse composite probabilities of two distributions with one fitted parameter.

What do these distributions mean? I have no adequate answer yet. For the present I would like to do some comments.

The distribution $W_1$ determines the energy levels of a particle in a centrosymmetric field ($\sim 1/r$), if its motion obeys quantum mechanics laws. Is it real quantization in gravitational field [4,5], or perhaps in electromagnetic one? H.Alfven [8] and F.Hoyle [9] were the first to discuss active influence of electromagnetic forces on the formation of the solar system. There is the remarkable fact that the total mechanic energy of hydrogen atom rotating around the sun on the orbit of Mercury $E_1 = -1.912 \cdot 10^{-18} J$ (see Table 1) is quite close to the energy of the first electronic level of hydrogen atom (ionization energy) $E_1 = -2.178 \cdot 10^{-18} J$. Is it chance coincidence?

The second distribution $W_2$ might be considered as Bose-Einstein energy distribution for spatial quantum oscillators provided chemical potential equals to zero and the exponent index equals to:

$$1.283n = \frac{1.283kT \cdot n}{kT} = \frac{\Delta E \cdot n}{kT} \qquad (7)$$

Where $\Delta E = 1.283kT$ is evidently the gap between equidistant energy levels.

The original equidistant energy levels with the gap of $\Delta E = 1.283kT$ are apparently manifested in the mass distribution in solar system. Suppose that the sun and planets were predominantly formed from identical particles. In this case the masses of the sun and planets must be proportional to the particle quantities containing in each of them. For energy levels of harmonic oscillator the logarithmic relationship of these quantities upon the level number must be a straight line with the slope equaled to $\Delta E/kT$. It can be seen from Fig. (3) that at least for the sun and major planets a straight line with the slope of 1.28 fits logarithm $M_n$ ($n = 0,6,7,8,9$) rather well *. The chemical potential in $W_2$ equaled to zero implies that condensation of the planets from the gaseous nebula took place at the temperature of the gas below the critical temperature, i.e. in the region of Bose-Einstein condensation. Therefore the solar system might be considered as Bose-Einstein condensate the vast majority of which falls on the basic energy level (the sun) and only the negligible quantity hits several first levels (the planets). It is not contradictory to the point that Bose-Einstein condensation is described as a "condensation in momentum space", because in gravitational field a spatial separation of two phases is liable to take place[10]. It is also known that the process of Bose-Einstein condensation is the first-order phase transition with the latent heat of transition be equaled to $\Delta Q = T\Delta S = 1.283kT$, where $T$ is the temperature of condensation and $\Delta S = 1.283k$ is the difference in entropy per particle between the gas phase and condensed phase[10]. It is highly plausible that this quantity determines the value of the gap between equidistant energy levels in Eq. (7).

Thus, if we suppose that initial nebular gas was degenerate then planetary orbits can be considered as discrete stationary one-particle energy levels with the distribution obtained in Eq. (4). If the Eq. (4) is true, then one or even several planets may exist in the solar system beyond Pluto. The parameters of the following planet are:
$a_{11} = 8.4 \cdot 10^{12} m$, $e_{11} \approx 0.18$, $T_{11} = 1.32 \cdot 10^{10} \sec \approx 420 y$, $M_{11} \approx 2 \cdot 10^{24} kg$.

___________

* As regards terrestrial planets and Asteroid Belt, the Gaussian distribution is suited for their mass allocation perfectly [16].

# APPENDIX

This paper was submitted to *Nature* on 30 January 2002 and was refused. Since that time a number of large trans-Neptunian objects (planetoids) were discovered in the Solar system (Table 2). Four former planetoids are undoubtedly the members of the classical Kuiper Belt, which is largely confined to heliocentric distances of 42 AU±10% [11,12]. 2003UB313 is probably a member of a scattered Kuiper Belt [13]. The discoverers of 2003VB12 consider its distant highly eccentric orbit as the result of scattering by a yet-to-be-discovered planet, or perturbation by an anomalously close stellar encounter [14].

As one can see none of parameters of discovered planetoids agree with predicted values of the new planet in this paper. However, they fit parameters of other predicted orbits well. The refined parameters of four predicted orbits beyond Neptune are listed in Table3.

It is currently suggested that not numerous dynamical family of Plutinos (the brightest object of which is Pluto) refers to inner region of Kuiper Belt. In this paper all the objects of the Main Asteroid Belt are held to occupy one orbit, namely, the fifth one. Now suppose it is accepted for all objects of Kuiper Belt (including Pluto), the mean heliocentric distance of which (42 AU) fits the semi-major axis of tenth orbit (Table 3: $a_{10}$ = 42,0 AU). On this assumption, the mass of Pluto needs to substitute for the total mass of Kuiper Belt in Fig. 3. The present total mass of Kuiper Belt is estimated up to several percent of Earth mass that is much more than for Pluto [11, 17, 18]. In the early stage it had to be still more massive, about ten Earth mass [11]. This value fits logarithmic mass – level number relationship (Fig. 3) much better, than the value of mass of Pluto.

According to Table 3, the object 2003UB313 (Xena) ranks twelfth, not eleventh, planet from the sun ($a_{12} = 66,7$ AU). If Xena possesses the Pluto's density (2000 kg/m$^3$) and the radius revealed (2800 km), then its mass is equal to approximately $2 \cdot 10^{23}$ kg. This value is consistent well with what is expected for the planet on twelfth orbit (see Table 3). By now there are no large TNO's, parameters of which are consistent with that of eleventh orbit. Therefore, one or more large planetoids having semi-major axes near 55 AU and a total mass a half Earth would be expected to discover.

Table 1. Observed and calculated characteristics of planetary orbits.

The values of semi-major axes ($a_n$) and eccentricities ($e_n$) were taken from [6]. Angular momentum ($L_n$) and total energy ($E_n^{ex}$) of the body on $n$-orbit have been calculated using Eq. (2) provided $m_n = m_H = 1.673 \cdot 10^{-27} kg$, $E_n^{th}$ - using Eq. (4) and $a_n^{th}$ - using Eq. (6).

| Planet | $n$ | $a_n$ | $e_n$ | $L_n$ | $E_n^{ex}$ | $E_n^{th}$ | $a_n^{th}$ |
|---|---|---|---|---|---|---|---|
| | | $\times 10^{-10}$, m | | $\times 10^{12}$, J·s | $\times 10^{18}$, J | $\times 10^{18}$, J | $\times 10^{-10}$, m |
| **Mercury** | 1 | 5.791 | 0.2056 | 4.54 | -1.918 | -1.874 | 5.92 |
| **Venus** | 2 | 10.821 | 0.00682 | 6.34 | -1.026 | -1.049 | 10.7 |
| **Earth** | 3 | 14.960 | 0.01675 | 7.46 | -0.742 | -0.763 | 14.5 |
| **Mars** | 4 | 22.79 | 0.09331 | 9.16 | -0.487 | -0.480 | 23.0 |
| **Ceres** | 5 | 41.39 | 0.079 | 12.4 | -0.268 | -0.268 | 41.3 |
| **Jupiter** | 6 | 77.83 | 0.04833 | 17.0 | -0.143 | -0.141 | 78.3 |
| **Saturn** | 7 | 142.8 | 0.05589 | 23.0 | -0.078 | -0.074 | 148 |
| **Uranus** | 8 | 287.2 | 0.0470 | 32.6 | -0.039 | -0.041 | 267 |
| **Neptune** | 9 | 449.8 | 0.0087 | 40.9 | -0.025 | -0.025 | 444 |
| **Pluto** | 10 | 591.0 | 0.247 | 45.4 | -0.019 | -0.018 | 629 |

Table 2. The parameters of newly discovered transneptunian objects (TNO's) and of Pluto.

| Data | Code name | Size, km | Mass, kg | Semi-major axis, AU | Semi-major axis, m | e |
|---|---|---|---|---|---|---|
| 1930 | Pluto | 2320 | $1,3·10^{22}$ | 39,3 | $5,9·10^{12}$ | 0,247 |
| June 2002 | 2003VB12 Quaoar | 1300 | | 43,4 | $6,5·10^{12}$ | 0,04 |
| December 2004 | 2003EL61 Santa | 2000-2500 | $4,2·10^{21}$ | 43,3 | $6,5·10^{12}$ | 0,19 |
| February 2004 | 2004DW | 1600 | | 45,0 | $6,7·10^{12}$ | |
| March 2005 | 2005FY9 Easterbunny | 1250 | | 45,6 | $6,8·10^{12}$ | 0,152 |
| October 2003 | 2003UB313 Xena | 2800 | | 67,71 | $10,2·10^{12}$ | 0,442 |
| November 2003 | 2003VB12 Sedna | 1500 | | 480 | $7,2·10^{13}$ | 0,84 |

Table 3. Parameters and populations (the mass of all objects which fall into the orbit) of four predicted planetary orbits following Neptune [this work and 15,16]

| Number of planetary orbit | Semi-major axis AU | Semi-major axis m | e | Mass, kg |
|---|---|---|---|---|
| 10 | 42,0 | $6,3·10^{12}$ | 0,30 | $~1,0·10^{25}$ |
| **11** | **55,3** | **$8,3·10^{12}$** | **0,38** | **$~2,8·10^{24}$** |
| 12 | 66,7 | $10,0·10^{12}$ | 0,41 | $~7,7·10^{23}$ |
| 13 | 82,7 | $12,4·10^{12}$ | 0,43 | $~2,1·10^{23}$ |

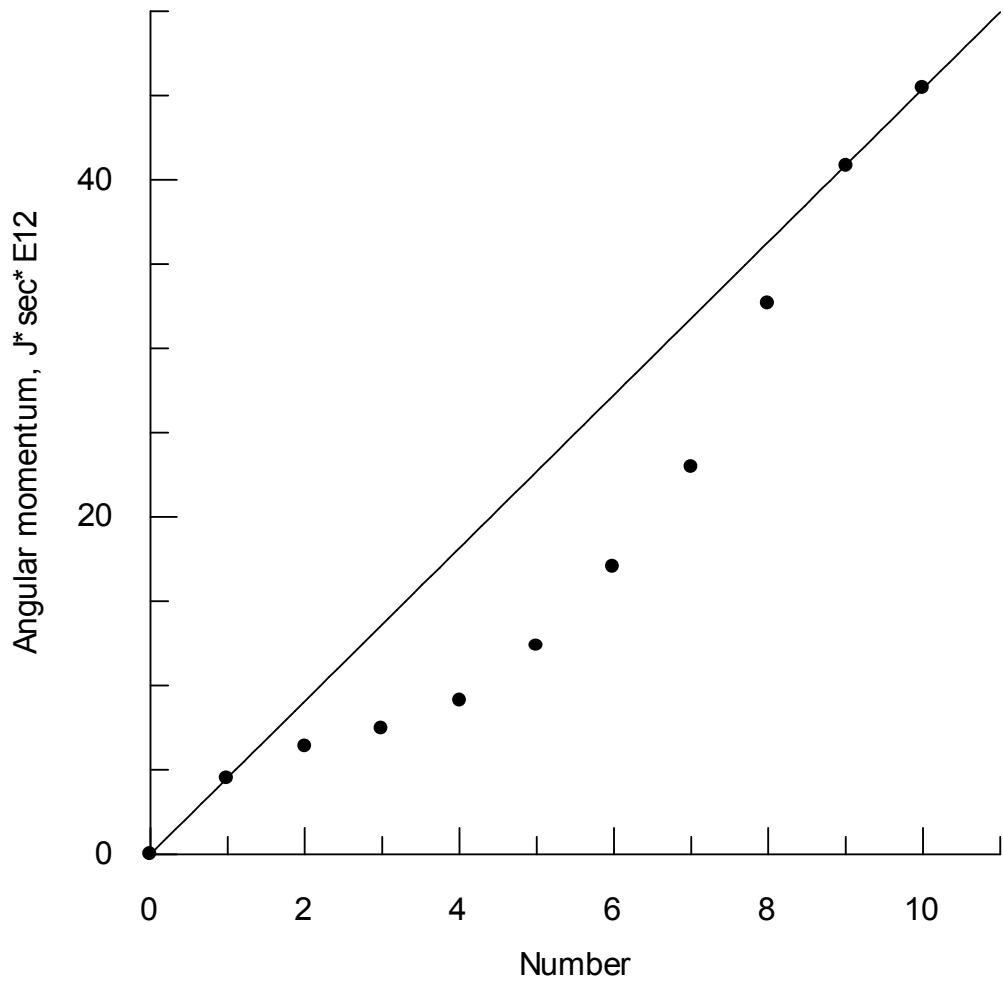

Fig.1 The planet's angular momentum per unit mass

( $m_n = m_H = 1.673 \cdot 10^{-27} kg$ ) vs. the corresponding number and a straight-line fit:
$L_n = 4.54 \cdot 10^{-12} n \ \ J \cdot \sec$.

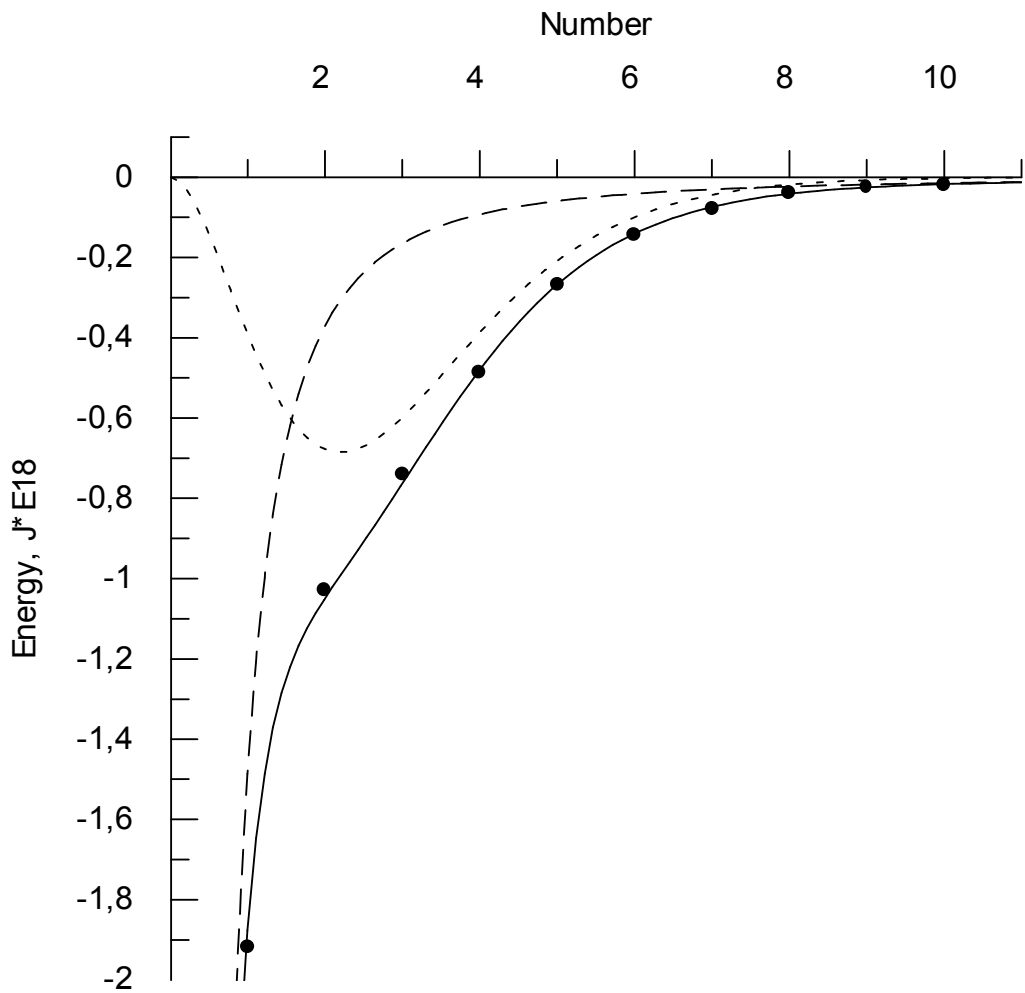

Fig.2 The energy profile of the solar system.

The black points are presented the values of $E_n^{ex}$ taken from Table 1 (6 column). The plain black line is simulated curve predicted by Eq. (4). The dashed line is the fraction of $W_1$, the dotted line is that of $W_2$ (see text).

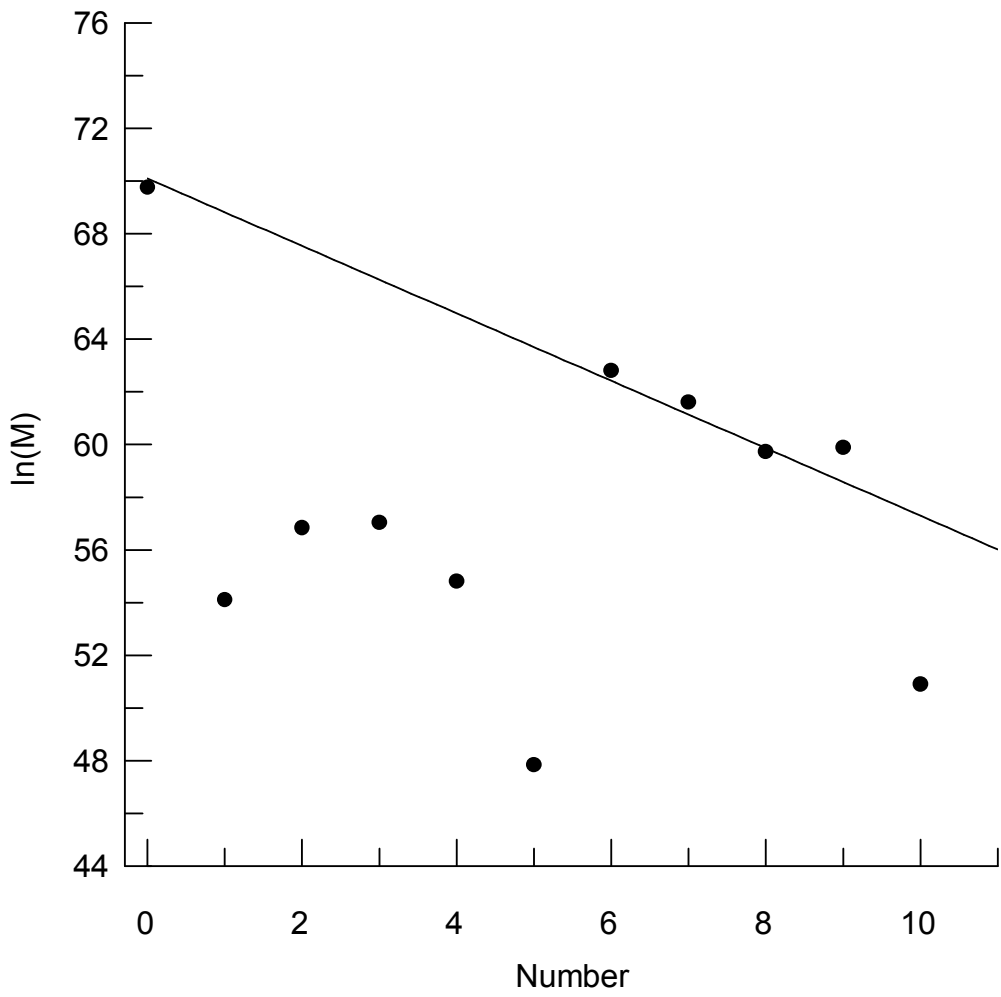

Fig.3 The semi-log plot of masses of the sun and planets vs. ordinal number and a straight-line fit $\ln(M_n) = 70.1 - 1.28n$.